\documentclass[conference]{IEEEtran}
\IEEEoverridecommandlockouts

\usepackage{cite}
\usepackage{amsmath,amssymb,amsfonts}
\usepackage{algorithmic}
\usepackage{graphicx}
\usepackage{textcomp}
\usepackage{xcolor}
\usepackage{url}
\usepackage{subcaption}
\usepackage{hyperref}
\usepackage{makecell}
\usepackage{tikz}
\usetikzlibrary{positioning, arrows.meta, calc, shadows}

\def\BibTeX{{\rm B\kern-.05em{\sc i\kern-.025em b}\kern-.08em
    T\kern-.1667em\lower.7ex\hbox{E}\kern-.125emX}}


\usepackage{fancyhdr}

\fancypagestyle{firstpage}{
    \fancyhf{}
    \fancyhead[C]{%
        \parbox{\textwidth}{%
            \centering
            \scriptsize
            © 2026 IEEE. Personal use of this material is permitted.
            Permission from IEEE must be obtained for all other uses, in any
            current or future media, including reprinting/republishing this
            material for advertising or promotional purposes, creating new
            collective works, for resale or redistribution to servers or lists,
            or reuse of any copyrighted component of this work.
        }%
    }
    
}

\begin{document}

\title{EmoNet: A Lightweight Attention-Based CNN\\for Speech Emotion Recognition}

\author{
\IEEEauthorblockN{Muhammad Moiz Khalid}
\IEEEauthorblockA{\textit{Department of Artificial Intelligence and Data Science} \\
\textit{National University of Computer and Emerging Sciences}\\
Islamabad, Pakistan \\
i232552@isb.nu.edu.pk}
\and
\IEEEauthorblockN{Abdur Rehman Tariq}
\IEEEauthorblockA{\textit{Department of Artificial Intelligence and Data Science} \\
\textit{National University of Computer and Emerging Sciences}\\
Islamabad, Pakistan \\
i232506@isb.nu.edu.pk}
\and
\IEEEauthorblockN{Muhammad Okasha Yamin}
\IEEEauthorblockA{\textit{Department of Artificial Intelligence and Data Science} \\
\textit{National University of Computer and Emerging Sciences}\\
Islamabad, Pakistan \\
i232525@isb.nu.edu.pk}
}

\maketitle
\thispagestyle{firstpage}
\begin{abstract}
Speech Emotion Recognition (SER) is a fundamental task in affective computing that enables applications such as mental health assessment, human--computer interaction, and automated customer support systems. Accurate emotion classification from speech remains challenging due to variations in speaker style, acoustic conditions, and the high computational cost of modern deep learning approaches. Many existing systems depend on large transformer-based architectures that require substantial computational resources, making them less suitable for real-time or resource-constrained environments. This research proposes EmoNet, a lightweight deep learning framework for speech emotion classification across seven emotion categories. The model is built on a four-block depthwise separable CNN encoder with squeeze-and-excitation attention and residual connections, combined with a Bidirectional LSTM for temporal feature learning and a multi-head attention pooling mechanism for improved representation. A comprehensive 134-channel acoustic feature set alongside data augmentation techniques including audio-level perturbations and SpecAugment is employed to improve robustness under diverse acoustic conditions. Evaluated on a combined dataset of TESS, RAVDESS, and SAVEE, EmoNet achieves a test accuracy of 96.75\% with only 4.11 million parameters. The proposed model outperforms its reproduced baseline and remains competitive with significantly larger state-of-the-art architectures.
\end{abstract}

\begin{IEEEkeywords}
EmoNet, Speech Emotion Recognition, Depthwise Separable Convolutions, Squeeze-and-Excitation, Bidirectional LSTM, Multi-Head Attention, Lightweight CNN, Multi-class Classification
\end{IEEEkeywords}

\section{INTRODUCTION}

Speech Emotion Recognition (SER) is the task of automatically identifying the emotional state expressed in a spoken utterance. Human speech contains rich paralinguistic information embedded in acoustic cues such as pitch, energy, tempo, and spectral characteristics. Emotional states influence nearly all aspects of the speech signal, making it challenging to learn representations that are both discriminative across emotion classes and robust to variations in speakers, recording conditions, and background noise \cite{barhoumi2026}. This balance between discriminability and robustness makes SER a complex problem in speech processing.

The importance of SER has increased considerably due to its wide range of real-world applications. In mental health monitoring, SER systems can assist in detecting early signs of stress, depression, or anxiety from conversational speech \cite{barhoumi2026}. In call-centre analytics, emotion-aware systems can improve customer experience through intelligent call routing and escalation handling. In automotive environments, SER can help detect driver fatigue, frustration, or panic, supporting safety-critical decisions. In social robotics and human--computer interaction, emotion recognition enables more adaptive and natural interactions. These applications have driven extensive research into deep learning-based SER systems.

Despite significant progress, deploying SER models in real-world environments remains challenging. Most existing approaches rely on large pre-trained transformer-based architectures such as Wav2Vec~2.0 \cite{baevski2020}, which contain hundreds of millions of parameters and require substantial computational resources, making them impractical for real-time or edge deployment. Another major challenge is limited generalization across datasets, as SER corpora are typically small and collected under different recording conditions. Models trained on a single dataset often fail to generalize to unseen domains, and combining multiple datasets introduces issues such as inconsistent labeling schemes, class imbalance, and domain shift. There is a strong need for lightweight yet accurate deep learning models that can classify speech emotion across multiple categories efficiently and in a way that can be applied in real-world systems.

In this paper, we propose EmoNet, a lightweight deep learning framework for speech emotion recognition that integrates a four-block depthwise separable CNN encoder with squeeze-and-excitation attention, a Bidirectional LSTM, and multi-head attention pooling to effectively capture both spatial and temporal dependencies in speech signals. In contrast to typical large transformer frameworks, EmoNet reduces computational complexity while maintaining competitive accuracy, enabling deployment in real-time and resource-constrained environments. EmoNet is trained and evaluated on a combined corpus of TESS, RAVDESS, and SAVEE covering seven emotion classes.

The main contributions of this paper are summarized as follows:
\begin{itemize}
    \item Proposing a lightweight SER architecture built on depthwise separable convolutions with squeeze-and-excitation channel attention, designed for efficient real-time emotion classification.
    \item Extending classification to seven emotion categories across a multi-corpus composite of TESS, RAVDESS, and SAVEE for improved generalization.
    \item Comparing performance to well-known deep learning models for speech emotion recognition.
\end{itemize}

The rest of this paper is organized as follows: Section~\ref{sec:literature} presents different related works in this field. Section~\ref{sec:method} outlines the proposed methodology. Experiments and results are shown in Section~\ref{sec:result}. Finally, Section~\ref{sec:conclusion} concludes the study.

\section{LITERATURE REVIEW}\label{sec:literature}

The potential of deep learning for automatic speech emotion recognition has attracted considerable research interest in recent years. A number of architectures and training strategies have been proposed to improve emotion classification performance, ranging from lightweight convolutional models to large pre-trained transformer systems. Here, we highlight significant recent research on the topic.

Chakrabarty \textit{et al.} \cite{chakrabarty2025} propose a hybrid Transformer-CNN framework that fuses a fine-tuned Wav2Vec 2.0 backbone with a 1D-CNN branch that contributes spectral stability through noise-resistant features including MFCCs, ZCR, and RMSE. An Attentive Temporal Pooling mechanism combines both streams, and the model is further equipped with an Explainable AI suite using SHAP and Score-CAM. The framework is validated across RAVDESS, TESS, SAVEE, and CREMA-D, achieving a near-perfect accuracy of 99.78\% on TESS. However, the Wav2Vec 2.0 backbone alone contributes approximately 95 million parameters \cite{baevski2020}, making the full model computationally prohibitive for resource-constrained or real-time deployment. Furthermore, the system evaluates individual corpora in isolation rather than a unified multi-corpus seven-class setting, limiting its applicability to cross-corpus generalization scenarios.

Wang \textit{et al.} \cite{wang2025} combine a fine-tuned Wav2Vec~2.0 encoder with a Neural Controlled Differential Equations (NCDE) classifier, evaluated on the four-class IEMOCAP dataset, attaining a weighted accuracy of 73.37\% and an unweighted accuracy of 74.18\% while converging within a single training epoch. Despite this convergence efficiency, the large parameter burden of Wav2Vec~2.0 and the constrained four-class evaluation limit direct comparability to broader seven-class benchmarks.

Kilimci \textit{et al.} \cite{kilimci2025} evaluate raw waveform-based deep learning frameworks for SER across six datasets, EMO-DB, RAVDESS, TESS, CREMA-D, SAVEE, and a combined TESS+RAVDESS set, without any handcrafted feature extraction stage. Multiple architectures are benchmarked, including SVMs, CNNs, LSTMs, and hybrid CNN-LSTM models. Their best result of 95.86\% is achieved by a CNN model on the combined TESS+RAVDESS set. While the raw waveform strategy achieves strong per-corpus results, the paper does not report a unified parameter count for the proposed models, making efficiency comparisons difficult. Importantly, datasets are evaluated separately rather than in a unified multi-corpus multi-class framework.

Li \cite{li2025} explores a CNN-Transformer hybrid framework for SER on a combined RAVDESS+SAVEE+TESS corpus, systematically investigating the impact of different loss functions on model accuracy and generalization. The system leverages Mel-Spectrogram features combined with multi-head self-attention to capture long-range temporal dependencies. While the combined dataset mirrors our experimental setup, the architecture does not employ lightweight depthwise separable convolutions, resulting in a heavier model. No explicit parameter count is provided, and the study focuses primarily on the loss function comparison rather than architectural efficiency.

Bhanbhro \textit{et al.} \cite{bhanbhro2025} present a comparative analysis of CNN-LSTM and attention-enhanced CNN-LSTM models for SER, evaluated primarily on the RAVDESS corpus. Their results consistently demonstrate that incorporating attention mechanisms into the CNN-LSTM pipeline improves classification accuracy and temporal modeling quality compared to vanilla CNN-LSTM baselines, validating the importance of attention for SER. However, the evaluation is limited to fewer than seven emotion classes and does not consider a multi-corpus composite training regime, which restricts the generalizability of conclusions that can be drawn from the study.

Barhoumi and BenAyed \cite{barhoumi2026} propose a transformer encoder combined with a comprehensive data augmentation pipeline, including noise injection, spectrogram shifting, and speed perturbation, to improve SER robustness. Both a Deep Neural Network and a Transformer Encoder are evaluated on RAVDESS, EMO-DB, and TESS, with the Transformer achieving 99.30\% on EMO-DB, 92.00\% on RAVDESS, and 100\% on TESS. These are strong individual-corpus results; however, the system is not evaluated in a composite multi-corpus regime and the transformer encoder architecture is not optimized for parameter efficiency, contrasting with the lightweight focus of our proposed system.

\section{PROPOSED METHODOLOGY}\label{sec:method}

As demonstrated in Section~\ref{sec:literature}, numerous studies have put forth different architectures and training strategies for SER. While some approaches achieved greater accuracy at the expense of intricate and computationally demanding models, particularly those relying on large pre-trained transformers, others reported less than ideal outcomes when confronted with multi-corpus or multi-class settings. Consequently, our goal in this paper is to enhance classification performance while reducing the amount of memory and processing power needed for computation.

\subsection{Datasets Employed}

EmoNet is trained and evaluated on a composite corpus assembled from three publicly available SER datasets. Table~\ref{tab:datasets} summarizes their properties. All audio is resampled to 16,000~Hz, padded or truncated to exactly 4~seconds (64,000 samples), and split in a stratified manner by class: 70\% training (3,304 samples), 15\% validation (708 samples), and 15\% testing (708 samples), using a fixed random seed of 42.

\begin{table}[h!]
\centering
\caption{Composite Corpus Statistics}
\label{tab:datasets}
\begin{tabular}{lcccc}
\hline
\textbf{Corpus} & \textbf{Speakers} & \textbf{Classes} & \textbf{Samples} & \textbf{Quality} \\
\hline
TESS \cite{dupuis2011}         & 2 (Female) & 7 & 2,800 & Studio       \\
RAVDESS \cite{livingstone2018} & 24 (Mixed) & 7 & 1,440 & Professional \\
SAVEE \cite{jackson2014}       & 4 (Male)   & 7 & 480   & Lab          \\
\hline
\textbf{Total} & \textbf{30} & \textbf{7} & \textbf{4,720} & -- \\
\hline
\end{tabular}
\end{table}

\textbf{TESS} (Toronto Emotional Speech Set) \cite{dupuis2011} contains 2,800 audio samples recorded by two female speakers from the University of Toronto, covering seven emotions: angry, disgust, fear, happy, neutral, surprise, and sad, with perfectly balanced classes of 400 samples each. The studio-controlled recording conditions make TESS a high-quality anchor corpus.

\textbf{RAVDESS} (Ryerson Audio-Visual Database of Emotional Speech and Song) \cite{livingstone2018} contributes 1,440 speech-only samples from 24 professional actors spanning both genders, covering all seven target emotions after mapping to the canonical taxonomy. RAVDESS adds substantial speaker diversity and varied vocal expression styles.

\textbf{SAVEE} (Surrey Audio-Visual Expressed Emotion) \cite{jackson2014} provides 480 samples from four male native British English speakers recorded at the University of Surrey. While the smallest corpus, SAVEE adds important acoustic diversity through a distinct accent and an exclusively male speaker pool, contrasting with the North American English of TESS and RAVDESS. 

\vspace{-5pt}
\subsection{Data Preprocessing \& Augmentation}\label{sec:augmentation}

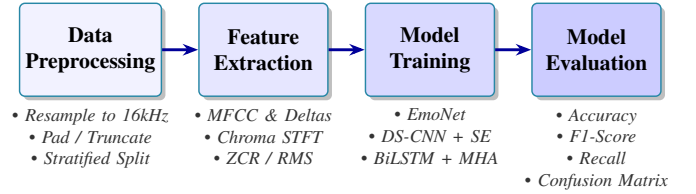
\begin{figure}[h!]
\centering
\resizebox{9cm}{!}{
\begin{tikzpicture}[
    node distance=0.7cm,
    base/.style={rectangle, rounded corners=3pt, draw=cyan!60!black, thick,
                 minimum width=2.5cm, minimum height=1.8cm, align=center,
                 fill=white, drop shadow, font=\large\bfseries},
    subtext/.style={font=\normalsize\itshape, text=black!80,
                    node distance=0.1cm, align=center},
    arrow/.style={-{Stealth[length=3mm]}, blue!60!black, ultra thick}
]
\node[base, fill=blue!5]  (prep)  {Data\\Preprocessing};
\node[subtext, below=of prep] (prep_d)
    {\textbullet~Resample to 16kHz\\ \textbullet~Pad / Truncate\\ \textbullet~Stratified Split};

\node[base, fill=blue!10, right=of prep] (feat) {Feature\\Extraction};
\node[subtext, below=of feat] (feat_d)
    {\textbullet~MFCC \& Deltas\\ \textbullet~Chroma STFT\\ \textbullet~ZCR / RMS};

\node[base, fill=blue!15, right=of feat] (train) {Model\\Training};
\node[subtext, below=of train] (train_d)
    {\textbullet~EmoNet\\ \textbullet~DS-CNN + SE\\ \textbullet~BiLSTM + MHA};

\node[base, fill=blue!20, right=of train] (eval) {Model\\Evaluation};
\node[subtext, below=of eval] (eval_d)
    {\textbullet~Accuracy\\ \textbullet~F1-Score\\ \textbullet~Recall\\ \textbullet~Confusion Matrix};

\draw[arrow] (prep)  -- (feat);
\draw[arrow] (feat)  -- (train);
\draw[arrow] (train) -- (eval);
\end{tikzpicture}
}
\caption{Proposed pipeline for Speech Emotion Classification}
\label{fig:pipeline}
\end{figure}

A two-stage augmentation pipeline is applied during training to improve generalization and robustness. At the audio level, each training sample is randomly subjected to one or two of: additive white Gaussian noise (SNR 10--25~dB), time stretching (rate 0.85--1.15), pitch shifting ($\pm$2 semitones), and random gain scaling ($-6$ to $+6$~dB). At the feature level, SpecAugment \cite{park2019} masks up to 25 time frames and 20 frequency bins per sample across two masks, forcing the model to learn robust representations from partial observations. The operational parameters for augmentation are summarized in Table~\ref{tab:augs}, and the overall processing pipeline is illustrated in Figure~\ref{fig:pipeline}.

\begin{table}[h!]
\centering
\caption{Data Augmentation Parameters}
\label{tab:augs}
\begin{tabular}{lc}
\hline
\textbf{Augmentation Parameter} & \textbf{Value} \\ \hline
Additive Noise (SNR)         & 10--25 dB                \\ \hline
Time Stretch Rate            & 0.85--1.15               \\ \hline
Pitch Shift                  & $\pm$2 semitones         \\ \hline
Random Gain Scaling          & $-6$ to $+6$ dB          \\ \hline
SpecAugment Time Masks       & Up to 25 frames (×2)     \\ \hline
SpecAugment Frequency Masks  & Up to 20 bins (×2)       \\ \hline
MixUp $\alpha$ (main phase)  & 0.3                      \\ \hline
MixUp $\alpha$ (fine-tune)   & 0.1                      \\ \hline
Resampling Rate              & 16,000 Hz                \\ \hline
Clip Duration                & 4 seconds (64,000 samples) \\ \hline
\end{tabular}
\end{table}

\subsection{Feature Extraction}

Rather than a simple MFCC vector, a 134-channel feature representation per time frame is extracted using a 512-point FFT and a 160-sample hop length:
\begin{equation}
\mathbf{f} = [\underbrace{\text{MFCC}}_{40},\;\underbrace{\Delta\text{MFCC}}_{40},\;\underbrace{\Delta^2\text{MFCC}}_{40},\;\underbrace{\text{Chroma}}_{12},\;\underbrace{\text{ZCR}}_{1},\;\underbrace{\text{RMS}}_{1}]
\tag{1}
\end{equation}
yielding $40+40+40+12+1+1=134$ channels. The 40 MFCCs capture the spectral shape of the audio. The 40 first-order delta coefficients capture the temporal rate of change of the spectral envelope, which is important for distinguishing emotions such as fear (rising pitch trajectory) from sadness (falling energy). The 40 delta-delta coefficients capture the second-order acceleration of spectral dynamics. The 12 chroma STFT features encode pitch class and harmonic content, while ZCR captures voicedness and RMS energy captures overall loudness. Each channel is normalized to zero mean and unit variance per sample to ensure equal contribution across features of different scales.

\subsection{Proposed Model Architecture}

\begin{figure}[h!]
    \centering
    \includegraphics[width=0.52\textwidth,keepaspectratio]{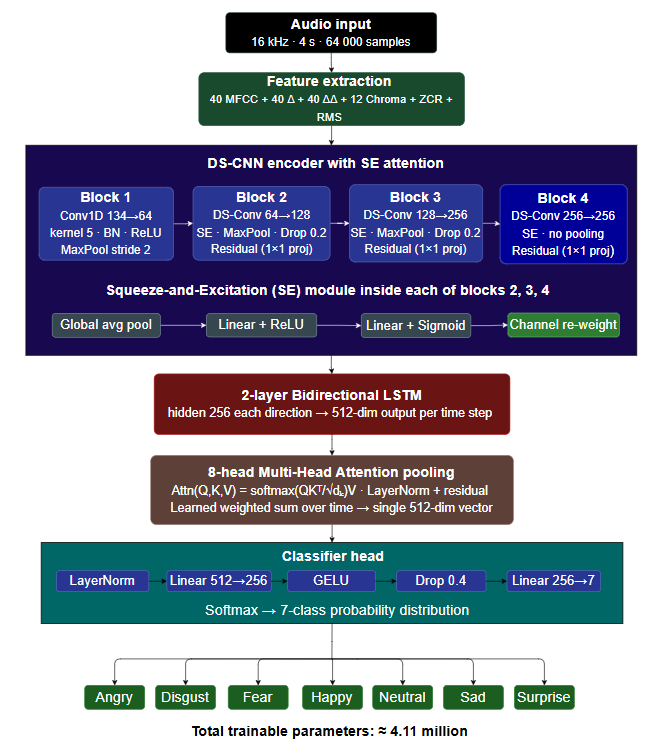}
    \caption{Proposed EmoNet Architecture}
    \label{fig:architecture}
\end{figure}

The model follows a four-stage pipeline, as depicted in Figure~\ref{fig:architecture}:
\[\small
\text{DS-CNN Encoder} \;\rightarrow\; \text{BiLSTM} \;\rightarrow\; \text{MHA Pooling} \;\rightarrow\; \text{Classifier}
\]
The total trainable parameter count is approximately 4.11 million, confirming that the system remains lightweight while capturing both spectral and temporal structure in speech. The hyperparameters used in training are shown in Table~\ref{tab:hyperparams}.

\subsubsection{DS-CNN Encoder}

The encoder comprises four sequential blocks. Block~1 applies a standard convolution (134$\to$64 channels, kernel size 5, padding 2), Batch Normalization, ReLU, and MaxPooling (stride 2). Blocks~2 and~3 each apply a Depthwise Separable Convolution with an integrated Squeeze-and-Excitation (SE) module, expanding 64$\to$128 and 128$\to$256 channels respectively, followed by MaxPooling and Dropout (0.2). Block~4 is an additional DS-Conv block (256$\to$256, Dropout 0.2) without pooling, preserving the temporal resolution of the feature sequence for the subsequent BiLSTM.

Each DS-Conv block factorizes a standard convolution into a depthwise spatial convolution followed by a pointwise $1{\times}1$ projection, substantially reducing the parameter count relative to full convolutions. The SE module uses global average pooling followed by two linear layers with a sigmoid gate to compute channel-wise attention weights, re-emphasizing informative spectral bands and suppressing less relevant ones:
\begin{equation}
s_c = \sigma\!\left(W_2\,\delta\!\left(W_1\,\frac{1}{T}\sum_{t=1}^{T}x_{c,t}\right)\right)
\tag{2}
\end{equation}
where $\delta$ is ReLU, $\sigma$ is sigmoid, and $T$ is the temporal dimension. Residual connections using $1{\times}1$ projection convolutions are applied in Blocks~2--4 to stabilize gradient flow, with adaptive average pooling used in the residual branch to match the pooled spatial dimensions.

\subsubsection{Bidirectional LSTM}

The CNN encoder output is passed to a 2-layer Bidirectional LSTM with hidden size 256 and inter-layer dropout of 0.3, yielding 512-dimensional bidirectional representations per time step. The BiLSTM captures long-range temporal dependencies that cannot be modeled by fixed-kernel convolutions, encoding how emotion unfolds dynamically across the utterance in both the forward and backward directions.

\subsubsection{Multi-Head Attention Pooling}

An 8-head Multi-Head Attention mechanism computes a learned weighted sum over the BiLSTM output sequence, concentrating on the most emotionally salient frames rather than uniformly averaging all time steps:
\begin{equation}
\text{Attn}(Q,K,V) = \text{softmax}\!\left(\frac{QK^\top}{\sqrt{d_k}}\right)V
\tag{3}
\end{equation}
Layer normalization and a residual connection are applied within the attention block. A linear scoring layer then produces scalar weights over time, which are passed through a softmax before computing the weighted sum over the attended representations.

\subsubsection{Classifier Head}

The pooled 512-dimensional representation passes through a series of fully connected layers: LayerNorm $\to$ Linear(512, 256) $\to$ GELU $\to$ Dropout(0.4) $\to$ Linear(256, 7), producing the final logits for classification. This combination of normalization, dimensionality reduction, non-linear activation, and dropout regularization creates a slim yet expressive classification head that retains strong prediction power while keeping the overall parameter count low.

\subsection{Training Recipe}

Training proceeds in two phases. The hyperparameters used are shown in Table~\ref{tab:hyperparams}.

\begin{table}[h!]
\centering
\caption{Hyperparameters used in EmoNet Model}
\label{tab:hyperparams}
\begin{tabular}{ll}
\hline
\textbf{Hyperparameter}            & \textbf{Value}                         \\ \hline
Optimizer                          & AdamW                                  \\ \hline
Learning Rate (main)               & $3\times10^{-4}$                       \\ \hline
Learning Rate (fine-tune)          & $5\times10^{-5}$                       \\ \hline
Weight Decay                       & $10^{-4}$                              \\ \hline
Batch Size                         & 32                                     \\ \hline
LR Schedule (main)                 & Cosine Annealing + Warm Restarts       \\ \hline
$T_0$, $T_\text{mult}$             & 20, 2                                  \\ \hline
LR Schedule (fine-tune)            & Cosine Annealing                       \\ \hline
MixUp $\alpha$ (main / fine-tune)  & 0.3 / 0.1                              \\ \hline
Label Smoothing (main / fine-tune) & 0.1 / 0.05                             \\ \hline
Dropout (main / fine-tune)         & 0.4 / 0.5                              \\ \hline
Gradient Clipping                  & 1.0                                    \\ \hline
Early Stopping Patience            & 15 epochs                              \\ \hline
\end{tabular}
\end{table}

\textbf{Main Phase.} AdamW \cite{loshchilov2019} is used with Cosine Annealing with Warm Restarts ($T_0=20$, $T_\text{mult}=2$, $\eta_\text{min}=10^{-6}$) for up to 80 epochs. MixUp \cite{zhang2018} is applied to 50\% of training batches ($\alpha=0.3$), creating convex label-interpolated combinations of sample pairs as a strong regularizer. Class-weighted cross-entropy with label smoothing of 0.1 handles class imbalance and prevents overconfidence. Gradient clipping at 1.0 prevents exploding gradients. Early stopping with a patience of 15 epochs is applied throughout.

\textbf{Fine-Tuning Phase.} The best main-phase checkpoint is reloaded; the learning rate is reduced to $5{\times}10^{-5}$ with a single-cycle Cosine Annealing schedule, label smoothing is reduced to 0.05, MixUp $\alpha$ is reduced to 0.1 (applied to 30\% of batches), and classifier Dropout is raised to 0.5. This phase sharpens decision boundaries without disrupting the generalizable representations formed during the main phase. After the model has been trained, the final component of the pipeline is evaluation.

All experiments were conducted on Google Colab using an NVIDIA T4 GPU, implemented in PyTorch 2.x with audio loading via \texttt{librosa}.

\section{RESULTS}\label{sec:result}

After training the model on the designated training set, its performance was assessed on a held-out test set consisting of 708 samples. Evaluation was conducted using several key metrics to measure the model's effectiveness. To apply these metrics accurately, it is crucial to define the following terms:

\begin{enumerate}
\item True Positive (TP): A positive instance correctly predicted as positive.
\item True Negative (TN): A negative instance correctly predicted as negative.
\item False Positive (FP): A negative instance incorrectly predicted as positive.
\item False Negative (FN): A positive instance incorrectly predicted as negative.
\item Accuracy: The proportion of instances correctly predicted by the model, reflecting its overall performance \cite{evidently_acc}.
\[
\text{Accuracy} = \frac{TP + TN}{TP + TN + FP + FN}
\tag{4}
\]
\item Confusion Matrix: A table that compares predicted labels with actual labels, showing the counts of true positives, true negatives, false positives, and false negatives \cite{evidently_cm}.
\item Precision: The proportion of correctly predicted positive instances among all predicted positives.
\[
\text{Precision} = \frac{TP}{TP + FP}
\tag{5}
\]
\item Recall: The proportion of actual positive instances correctly identified by the model \cite{evidently_acc}.
\[
\text{Recall} = \frac{TP}{TP + FN}
\tag{6}
\]
\item F1 Score: The harmonic mean of precision and recall, providing a balanced assessment of model performance by accounting for both false positives and false negatives.
\[
\text{F1 Score} = \frac{2 \times \text{Precision} \times \text{Recall}}{\text{Precision} + \text{Recall}}
\tag{7}
\]
\end{enumerate}

These metrics collectively provide a nuanced understanding of the model's performance, highlighting its strengths and areas for improvement.

\begin{figure}[h!]
    \centering
    \includegraphics[width=1.0\linewidth]{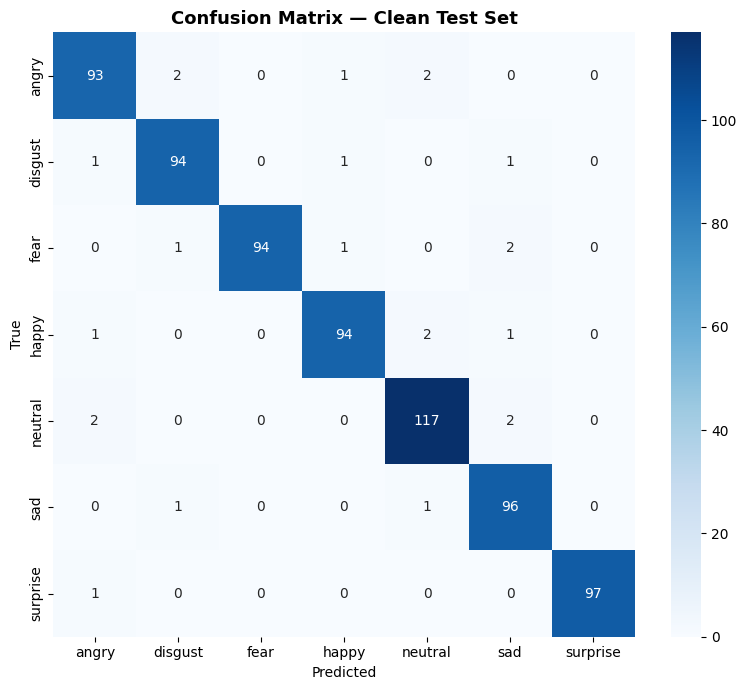}
    \caption{Confusion Matrix}
    \label{fig:conf_matrix}
\end{figure}

During the building phase of our model's architecture, we thoroughly tested many different configurations and hyperparameters in order to find the best combination. The final EmoNet model achieved a test accuracy of 96.75\% with only 4.11 million parameters, outperforming the reproduced baseline while using approximately 34$\times$ fewer parameters. The performance metrics of the model are displayed in Table~\ref{tab:performance_metrics}, and further details are shown in Figure~\ref{fig:conf_matrix}.

\begin{table}[h!]
\centering
\caption{Overall Test Set Performance}
\label{tab:performance_metrics}
\begin{tabular}{ll}
\hline
\textbf{Metric}            & \textbf{EmoNet} \\ \hline
Accuracy (ACC)             & 96.75\%         \\ \hline
Weighted Precision         & 96.79\%         \\ \hline
Weighted Recall            & 96.75\%         \\ \hline
Weighted F1-Score          & 96.76\%         \\ \hline
Total Parameters           & 4.11M           \\ \hline
\end{tabular}
\end{table}

The strong performance of EmoNet is supported by several key design choices. The depthwise separable CNN encoder with squeeze-and-excitation attention provides effective spectral feature extraction while keeping the parameter count low. The Bidirectional LSTM captures how emotion unfolds temporally across an utterance, which fixed-kernel convolutions alone cannot model. The multi-head attention pooling concentrates on the most emotionally salient frames rather than averaging uniformly across all time steps. The fourth CNN block, which operates without pooling, preserves temporal resolution and provides the BiLSTM with a longer and richer sequence to model long-range dependencies. Together, these components enable EmoNet to achieve high accuracy with far fewer parameters than conventional transformer-based models.

\begin{table}[h!]
\centering
\caption{Per-Class Performance on Held-Out Test Set}
\label{tab:perclass}
\begin{tabular}{lcccc}
\hline
\textbf{Emotion} & \textbf{Precision} & \textbf{Recall} & \textbf{F1} & \textbf{Support} \\ \hline
Angry    & 0.95 & 0.95 & 0.95 & 98  \\
Disgust  & 0.96 & 0.97 & 0.96 & 97  \\
Fear     & 1.00 & 0.96 & 0.98 & 98  \\
Happy    & 0.97 & 0.96 & 0.96 & 98  \\
Neutral  & 0.96 & 0.97 & 0.96 & 121 \\
Sad      & 0.94 & 0.98 & 0.96 & 98  \\
Surprise & 1.00 & 0.99 & 0.99 & 98  \\ \hline
\textbf{Weighted Avg} & \textbf{0.97} & \textbf{0.97} & \textbf{0.97} & 708 \\ \hline
\end{tabular}
\end{table}

All seven classes achieve F1 scores between 0.95 and 0.99, as shown in Table~\ref{tab:perclass}. The \textit{surprise} class achieves the highest F1 of 0.99, benefiting from clean and consistent recordings in both TESS and RAVDESS. The \textit{sad} class shows the lowest precision of 0.94, reflecting its acoustic overlap with neutral and disgust; all three emotions are characterized by relatively low energy and slow speaking tempo.

Table~\ref{tab:baseline_comparison} compares EmoNet against the reproduced baseline. Our model achieves higher performance across all evaluation metrics, with accuracy improving from 92.66\% to 96.75\%. Moreover, the reproduced model contains 139.24M parameters, whereas EmoNet uses only 4.11M, demonstrating that the proposed approach is more computationally efficient while achieving superior predictive performance.

\begin{table}[h!]
\centering
\caption{Comparison with Reproduced Baseline}
\label{tab:baseline_comparison}
\begin{tabular}{|l|c|c|c|}
\hline
\textbf{Metric} & \textbf{Paper Results} & \textbf{Reproduced} & \textbf{EmoNet (Ours)} \\ \hline
Accuracy (\%)   & 98.91 & 92.66 & 96.75 \\ \hline
Precision       & 0.977 & 0.927 & 0.968 \\ \hline
Recall          & 0.977 & 0.927 & 0.968 \\ \hline
F1-Score        & 0.977 & 0.926 & 0.968 \\ \hline
Params (M)      & 140   & 139.24 & 4.11  \\ \hline
\end{tabular}
\end{table}

Table~\ref{tab:corpus} reports results evaluated separately on the test subset of each source corpus to assess cross-domain generalization. TESS achieves perfect accuracy, reflecting its studio-controlled conditions. RAVDESS reaches 92.73\%, a strong result for professional actor speech across 24 diverse speakers. SAVEE achieves 90.79\% despite being the smallest corpus and representing a distinct British English accent and an exclusively male speaker demographic, confirming that EmoNet generalizes well across different acoustic conditions and speaker characteristics.

\begin{table}[h!]
\centering
\caption{Per-Corpus Test Accuracy and Weighted F1}
\label{tab:corpus}
\begin{tabular}{lccc}
\hline
\textbf{Corpus} & \textbf{Accuracy} & \textbf{F1 (Weighted)} & \textbf{N} \\ \hline
TESS    & 1.0000 & 1.0000 & 412 \\
RAVDESS & 0.9273 & 0.9274 & 220 \\
SAVEE   & 0.9079 & 0.9077 & 76  \\ \hline
\end{tabular}
\end{table}

\section{CONCLUSION}\label{sec:conclusion}

In this study, we introduced EmoNet, a lightweight speech emotion recognition framework built on depthwise separable convolutional layers augmented with squeeze-and-excitation channel attention, a two-layer Bidirectional LSTM, and 8-head multi-head attention pooling. The EmoNet architecture was designed to improve upon large transformer-based baselines by drastically lowering the total parameter count while maintaining strong predictive capacity. Trained on a 4,720-sample composite of TESS, RAVDESS, and SAVEE using a two-phase training strategy incorporating MixUp, SpecAugment, and Cosine Annealing with Warm Restarts, EmoNet achieves a 96.75\% test accuracy across seven emotion classes with only 4.11 million parameters, rivaling state-of-the-art models with significantly more parameters, as summarized in Table~\ref{tab:comparison} and reported in Section~\ref{sec:literature}. Comparisons with recent approaches show that EmoNet strikes an effective balance between accuracy and computational efficiency, making it well-suited for real-time and resource-constrained deployment.

\begin{table}[h!]
\centering
\caption{Model Comparison with Related Work}
\label{tab:comparison}
\scriptsize
\setlength{\tabcolsep}{1.5pt}
\begin{tabular}{|l|c|c|c|c|}
\hline
\textbf{Model} & \textbf{Acc. (\%)} & \textbf{Classes} & \textbf{Params} & \textbf{Corpus} \\ \hline
Chakrabarty \textit{et al.} \cite{chakrabarty2025} & 99.78 & 7  & $\sim$95M & TESS (single) \\ \hline
Kilimci \textit{et al.} \cite{kilimci2025} & 95.86 & -- & N/R & TESS+RAVDESS \\ \hline
Barhoumi \& BenAyed \cite{barhoumi2026} & 100 & 7 & N/R & TESS (single) \\ \hline
Li \cite{li2025} & -- & 7 & N/R & RAVDESS+SAVEE+TESS \\ \hline
\textbf{EmoNet (Ours)} & \textbf{96.75} & \textbf{7} & \textbf{4.11M} & \textbf{TESS+RAVDESS+SAVEE} \\ \hline
\end{tabular}
\end{table}

Going forward, future improvements may take the approach of including naturalistic and non-acted speech corpora such as CREMA-D and IEMOCAP to better assess real-world robustness. Testing the model on external datasets may further examine its generalization and reliability for real-world clinical and commercial deployment. Such evaluation would also help assess how well the model performs on spontaneous conversational speech, which may differ substantially from the acted, studio-quality recordings used in this study. Moreover, examining various other lightweight architectures, extending the augmentation pipeline, and systematic hyperparameter tuning could further enhance model performance.


\begin{thebibliography}{00}

\bibitem{barhoumi2026}
C.~Barhoumi and Y.~BenAyed, ``Transformer Encoder and Data Augmentation for Real-Time Speech Emotion Recognition,'' \textit{Multimedia Tools and Applications}, vol.~85, article no.~226, 2026.

\bibitem{baevski2020}
A.~Baevski, Y.~Zhou, A.~Mohamed, and M.~Auli, ``Wav2vec 2.0: A Framework for Self-Supervised Learning of Speech Representations,'' \textit{Advances in Neural Information Processing Systems}, vol.~33, pp.~12449--12460, 2020.

\bibitem{chakrabarty2025}
S.~Chakrabarty, P.~Bishwas, and R.~Chatterjee, ``Explainable Transformer-CNN Fusion for Noise-Robust Speech Emotion Recognition,'' \textit{arXiv preprint arXiv:2512.18298}, Dec.\ 2025.

\bibitem{wang2025}
N.~Wang \textit{et al.}, ``Speech Emotion Recognition Using Fine-Tuned Wav2vec2.0 and Neural Controlled Differential Equations Classifier,'' \textit{PLOS ONE}, vol.~20, no.~2, p.~e0318297, Feb.\ 2025.

\bibitem{kilimci2025}
Z.~H.~Kilimci, {\"U}.~Bayraktar, and A.~K{\"u}{\c{c}}{\"u}kmanisa, ``Evaluating Raw Waveforms with Deep Learning Frameworks for Speech Emotion Recognition,'' \textit{Multimedia Tools and Applications}, vol.~84, pp.~45119--45149, 2025.

\bibitem{li2025}
B.~Li, ``Speech Emotion Recognition Based on CNN-Transformer with Different Loss Function,'' \textit{Journal of Computer and Communications}, vol.~13, no.~3, pp.~103--115, 2025.

\bibitem{bhanbhro2025}
J.~Bhanbhro \textit{et al.}, ``Speech Emotion Recognition: Comparative Analysis of CNN-LSTM and Attention-Enhanced CNN-LSTM Models,'' \textit{Signals}, vol.~6, no.~2, p.~22, 2025.

\bibitem{dupuis2011}Toronto emotional speech set (TESS). (2019). Kaggle. \href{https://www.kaggle.com/datasets/ejlok1/toronto-emotional-speech-set-tess}{https://www.kaggle.com/datasets/ejlok1/toronto-emotional-speech-set-tess}

\bibitem{livingstone2018}RAVDESS emotional speech audio. (2019). Kaggle. \href{https://www.kaggle.com/datasets/uwrfkaggler/ravdess-emotional-speech-audio}{https://www.kaggle.com/datasets/uwrfkaggler/ravdess-emotional-speech-audio}

\bibitem{jackson2014}Surrey audiovisual expressed emotion (SAVEE). (2019). Kaggle. \href{https://www.kaggle.com/datasets/ejlok1/surrey-audiovisual-expressed-emotion-savee}{https://www.kaggle.com/datasets/ejlok1/surrey-audiovisual-expressed-emotion-savee}

\bibitem{park2019}
D.~S.~Park \textit{et al.}, ``SpecAugment: A Simple Data Augmentation Method for Automatic Speech Recognition,'' in \textit{Proc.\ Interspeech}, 2019, pp.~2613--2617.

\bibitem{loshchilov2019}
I.~Loshchilov and F.~Hutter, ``Decoupled Weight Decay Regularization,'' in \textit{Proc.\ ICLR}, 2019.

\bibitem{zhang2018}
H.~Zhang, M.~Ciss{\'e}, Y.~N.~Dauphin, and D.~Lopez-Paz, ``MixUp: Beyond Empirical Risk Minimization,'' in \textit{Proc.\ ICLR}, 2018.

\bibitem{evidently_acc}
Evidently AI, ``Classification metrics guide: Accuracy vs. precision vs. recall in machine learning: what's the difference?,'' 2024. [Online]. Available: \url{https://www.evidentlyai.com/classification-metrics/accuracy-precision-recall}

\bibitem{evidently_cm}
Evidently AI, ``Classification metrics guide: How to interpret a confusion matrix for a machine learning model,'' 2024. [Online]. Available: \url{https://www.evidentlyai.com/classification-metrics/confusion-matrix}

\end{thebibliography}
\end{document}